\documentclass[useAMS,usenatbib]{mn2e}

\usepackage{widetext}
\usepackage{eucal}
\usepackage{aas_macros}
\usepackage{amsmath}
\usepackage{amssymb}
\usepackage{epsfig}
\usepackage{graphicx}
\usepackage{ifthen}
\usepackage{latexsym}
\usepackage{ogonek}
\usepackage{rotating}
\usepackage{subfigure}
\usepackage{times,epsf}
\usepackage{txfonts}
\usepackage{varioref}
\usepackage{verbatim}
\usepackage{url}
\numberwithin{equation}{section}  

\newcommand{\beq}{\begin{equation}}
\newcommand{\eeq}{\end{equation}}

\newcommand{\abr}{\mathcal{L}_*}
\newcommand{\shear}{\sigma_{\hat{r}\hat{\phi}}}

\newcommand{\curly}[1]{\mathcal{#1}}
\newcommand{\astar}{a_*}
\newcommand{\mstar}{M_{*}}
\newcommand{\mdot}{\dot{M}}
\newcommand{\mdotstar}{\dot{M}_*}
\newcommand{\opacity}{\bar{\kappa}}
\newcommand{\vr}{v^{\hat{r}}}

\begin{document}

\title[Thin Disk Theory with a Non-Zero Boundary Torque]{Thin Disk Theory with a Non-Zero Torque
 Boundary Condition and Comparisons with Simulations}

\author[R.~F. Penna, A.~S. S\k{a}dowski, \& J.~C. McKinney]
{
Robert F. Penna$^1$\thanks{E-mail: 
 rpenna@cfa.harvard.edu~(RFP),
 asadowski@cfa.harvard.edu~(AS),
 jmckinne@stanford.edu~(JCM),},
 Aleksander S\k{a}dowski$^1$\footnotemark[1],
 Jonathan C. McKinney$^2$$^3$\footnotemark[1]\\
  $^1$Harvard-Smithsonian Center for Astrophysics, 60 Garden Street, Cambridge, MA 02138, USA \\
  $^2$Department of Physics and Kavli Institute for Particle Astrophysics and Cosmology, Stanford University, Stanford, CA 94305-4060, USA \\
  $^3$Chandra Fellow \\
}
\date{\today}

\maketitle

\begin{abstract}

We present an analytical solution for thin disk accretion onto a Kerr
black hole that extends the standard Novikov-Thorne $\alpha$-disk in
three ways: (i) it incorporates nonzero stresses at the inner edge of
the disk, (ii) it extends into the plunging region, and (iii) it uses
a corrected vertical gravity formula.  The free parameters of the
model are unchanged.  Nonzero boundary stresses are included by
replacing the Novikov-Thorne no torque boundary condition with the
less strict requirement that the fluid velocity at the innermost
stable circular orbit is the sound speed, which numerical models show
to be the correct behavior for luminosities below $\sim 30\%$
Eddington.  We assume the disk is thin so we can ignore advection.
Boundary stresses scale as $\alpha h$ and advection terms scale as
$h^2$ (where $h$ is the disk opening angle $(h=H/r)$), so the model is
self-consistent when $h < \alpha$.  We compare our solution with slim
disk models and general relativistic magnetohydrodynamic disk
simulations.  The model may improve the accuracy of black hole spin
measurements.

\end{abstract}


\section{Introduction}
\label{sec:intro}

The standard model for relativistic, thin disk accretion onto a black
hole is the Novikov-Thorne (NT) model \citep{nt73,pt74}.  It is the
relativistic generalization of the \citet{sha73} $\alpha$-disk.  The
black hole is described by the Kerr metric with fixed mass, $M$, and
specific angular momentum, $a$.  The accretion flow is razor thin and
confined to the equatorial plane so heat advection is negligible and
vertical and radial energy transport can be decoupled.  The disk has an
inner edge at the innermost stable circular orbit (ISCO) where viscous
stresses are assumed to vanish.  
Mass is accreted at a rate $\dot{M}$ as stresses and radiation
transport energy and angular momentum outwards.  The NT disk has four
free parameters: $M$, $a$, $\dot{M}$, and $\alpha$.

Slim disk models generalize the NT model to include heat advection and
coupled radial and vertical energy transport. They are solved
numerically by requiring the flow to pass smoothly through a sonic
point
\citep{abr88,pac00,ap03,snm08,sadowski_slim_disks_2009,abr10,sadowski11}.
They have the same free parameters as the NT disk.

General relativistic, magnetohydrodynamic (GRMHD) thin disk
simulations \citep{shafee08,noblekrolik09,nkh10,rfp10} incorporate
magnetic fields and turbulence is driven by the magnetorotational
instability (MRI) \citep{bal91,bh98}.  Radiation physics is described
in an ad-hoc way as a sink term in the energy equation that tends to
drive the entropy of the fluid towards a target, $K_c$.  The MRI
generates turbulence self-consistently, so the $\alpha$-viscosity
prescription is not used.  However the strength of the saturated magnetic
stresses, the ``effective $\alpha$,'' depends on the gas-to-magnetic
pressure ratio, $\beta_i$, of the initial fields in the
simulation. 
So the target entropy, $K_c$, plays a similar role to
$\dot{M}$, and $\beta_i$ is analogous to $\alpha$.

\citet{nkh10} presented simulations of GRMHD disks with different
thicknesses and compared their stress profiles.  They argued there is
a large stress at the inner edge even in the limit of vanishing disk
opening angle $h\rightarrow 0$.
They concluded magnetized disks cannot be described by the NT
model independently of disk thickness.  \citet{rfp10} performed
similar simulations but found the stress at the inner edge to be
directly proportional to thickness.  They argued the NT zero-stress
boundary condition is valid in the limit $h\rightarrow
0$.  We believe the difference is in their definitions of the stress.
\citet{nkh10} included all of the fluid.  \citet{rfp10} made a
distinction between the high density disk region and the low density,
highly magnetized, coronal region.  They included only disk fluid in
stress calculations. The corona, if it is included, makes a
large contribution to the stress.  

The present model only describes the high-density disk region, which
is expected to dominate the emission leading to the observed thermal
spectral component.  The corona is expected to contribute mostly to
non-thermal spectral components.  So the \citet{rfp10} result that
stress scales with $h$ is the relevant one.

Our model includes a nonzero stress at the inner edge of the disk.
Nonzero stress boundary conditions have been previously considered in
the context of Newtonian \citep{shapirobook83} and general
relativistic \citep{ak00} accretion.  However, the stress at the inner
edge is a free parameter in these models.  We eliminate this parameter
by identifying the inner edge with the sonic point and relating the
stress there to the sound speed.  This prescription reduces to the NT
zero-torque boundary condition in the razor thin limit
$h \rightarrow 0$.

In the next section we describe the differences between our model and
NT.  In \S\ref{sec:gtd} we give the explicit disk solution.  In
\S\ref{sec:slim} we compare it to slim disk models and in
\S\ref{sec:grmhd} we compare it to GRMHD disk simulations.  We
summarize our main results in \S\ref{sec:conclude}.  The Kerr metric
and the disk structure equations are summarized in the appendices.  A
Fortran code which computes our thin disk solutions is available at
{\tt https://www.cfa.harvard.edu/$\sim$rpenna/thindisk}.

\section{Physics beyond the standard disk model}
\label{sec:approx}

Our model extends the NT model in three ways: (i) it incorporates
nonzero stresses at the inner edge of the disk, (ii) it extends into
the plunging region, and (iii) it uses the correct vertical gravity
formula.  In this section we discuss each of these developments.

\subsection{The inner edge boundary condition}
\label{sec:bc}

\begin{figure}
\includegraphics[width=3.3in,clip]{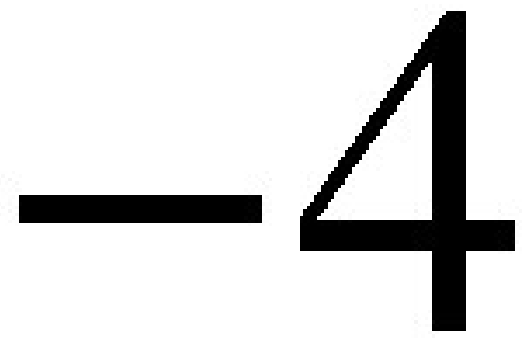}
\caption{Time-averaged rest mass density in the $r-z$ plane for four
  GRMHD simulations with $a_*=0$ and various disk thicknesses.  The
  dashed vertical line marks the ISCO.  The disk opening angle, $h$,
  and effective Shakura-Sunyaev viscosity, $\alpha$, are measured at
  the sonic point, $r_0$ (c.f. \S\ref{sec:grmhd}).  The top three
  panels have $h \ll \alpha$ and the inner edge of the disk is located
  outside the ISCO.  The lowermost panel has $h \gg \alpha$ and the
  density increases monotonically down to the event horizon.  Panels
  (a) and (b) are the thin and thick $a_*=0$ simulations of
  \citet{kulkarni11}.  Panels (c) and (d) are models A0HR2 and A0HR3,
  respectively, from \citet{rfp10}.}
\label{fig:densitya0}
\end{figure}

\begin{figure}
\includegraphics[width=3.3in,clip]{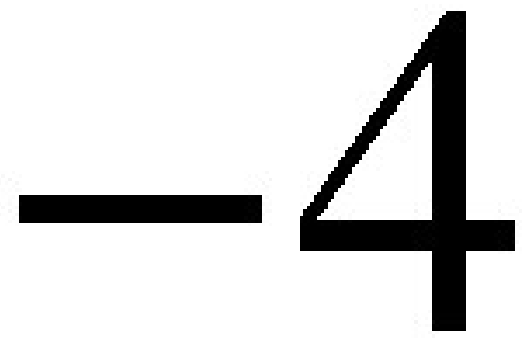}
\caption{Same as Figure \ref{fig:densitya0} but with $a_*=0.7$.  Panel
(a) is the $a_*=0.7$ simulation of \citet{kulkarni11}.  Panels (b),
(c), and (d) are models A7HR1, A7HR2, and A7HR3, respectively, from
\citet{rfp10}.}
\label{fig:densitya7}
\end{figure}

The criterion $h \ll \alpha$, where $h$ is the disk opening angle $(h=H/r)$ and
$\alpha$ is the ``effective viscosity'' parameter
(c.f. \S\ref{sec:grmhd}), governs the structure of weakly-magnetized
GRMHD disk simulations.  When the disk is thin, $h \ll \alpha$, the
surface density has an inner edge near the sonic point.  When the disk
is thick, $h \gg \alpha$, advection causes the disk density to increase
monotonically down to the event horizon.  This is illustrated in
Figures \ref{fig:densitya0} and \ref{fig:densitya7}, which show
time-averaged rest mass density in the $r-z$ plane for eight GRMHD
simulations. Rest mass density drops as the disk approaches
the sonic point if and only if $h \ll \alpha$.

Thick disks are non-Keplerian and insensitive to the ISCO.  The sonic
point and ISCO only coincide if the disk is thin.  \citet{abr10} find
the Boyer-Lindquist radial positions of the sonic point and ISCO
deviate by $< 3\%$, independently of $\alpha$, for $a_*=0$ slim disks
with $\dot{M}/\dot{M}_{\rm edd}<0.3$.

Summarizing these two observations: the inner edge, sonic point, and
ISCO are at the same radius if $h \ll \alpha$ and
$\dot{M}/\dot{M}_{\rm edd}<0.3$.  We will assume these conditions hold for
thin disks.  Under these conditions, energy advection and energy
generation by compression, which scale as $h^2$, can be neglected
relative to boundary stresses at the ISCO, which scale as $\alpha h$.
We give a proof in Appendix \ref{app:bc}.  The NT model
ignores boundary stresses at the ISCO, which is valid in the limit
$h\rightarrow 0$.  By including them, we obtain disk solutions with
nonzero stress and flux at the ISCO, and radial velocity and surface
density profiles that can be extended into the plunging region.

\subsection{The plunging region}
\label{sec:plunge}

We assume magnetic fields are weak and plunging region stresses are
small, so the fluid motion inside the ISCO can be approximated by
geodesics.  \citet{krolik99} argued magnetic fields are always
dynamically important inside the ISCO and geodesic trajectories are
never a good approximation there.  However we will show in
\S\ref{sec:grmhd} that our solution gives a good fit to the radial
velocity profile of the fiducial $\astar=0$ GRMHD simulation of
\citet{kulkarni11}.  So GRMHD disks do exist in which the field is
sufficiently weak inside the ISCO that geodesic motion is a good
approximation.

To solve the geodesic equations (\ref{eq:tdot})-(\ref{eq:phidot}) for
the fluid motion, we need to fix the fluid energy $E=|u_t|$ and
angular momentum $L=u_\phi$.  We assume the fluid plunges with
constant $E$ and $L$, so it suffices to fix the fluid energy and
angular momentum at the ISCO.  Equivalently, it is enough to specify the
fluid velocity at the ISCO.

We assume the angular velocity at the ISCO is Keplerian
(\ref{eq:omega}) and the disk is in the equatorial plane, so
$u^\theta=0$.  The radial velocity at
the ISCO is the sound speed (c.f. \S\ref{sec:bc}):
\begin{equation}
V_{\rm 0}=\left(h \frac{\abr}{r}\right)_0.
\end{equation}
This is the radial velocity in the fluid frame.  It is related to
$u^r$ by
$V=\sqrt{g_{rr}}u^r/\left(1+\left(\sqrt{g_{rr}}u^r\right)^2\right)^{1/2}$.

The flow inside the ISCO is now fixed by the disk structure
equations.  See Appendix \ref{app:disk} for details.

\subsection{Vertical gravity}
\label{sec:vertg}

NT gave an incorrect formula for the ``vertical gravity'' appearing in
the pressure balance equation, creating errors in the disk solution.
Better formulae for the vertical gravity were found by
\citet{eardley75} and \citet{rif95}.  However they assume the disk follows
circular geodesics, so their results break down in the plunging
region.  \citet{abr97} found a more general formula which is valid in
the plunging region.  We use this result in the disk structure
equations (\ref{eq:hydroeq}).

\begin{equation*}
\\
\\
\end{equation*}

\section{Explicit disk solutions}
\label{sec:gtd}

The model has four free parameters:
\begin{align*}
M      &= \text{mass of black hole},\\
a      &= \text{specific angular momentum of hole},\\
\mdot  &= \text{accretion rate},\\
\alpha &= \text{Shakura-Sunyaev viscosity}.
\end{align*}
These are taken to be constants.  Following NT, we shall express $M$
in units of $3 M_\odot$ and we shall express $\mdot$ in units of
$10^{17} \text{ g/sec}$:
\begin{equation*}
\mstar    \equiv M/3 M_\odot,
\quad
\mdotstar \equiv \mdot /10^{17} \text{ g sec$^{-1}$}.
\end{equation*}

The metric and disk structure equations are summarized in Appendices
\ref{app:metric} and \ref{app:disk}.  In this section we give the
solutions for the quantities that appear in the structure equations\footnote{A
Fortran code which computes our thin disk solutions is available at
{\tt https://www.cfa.harvard.edu/$\sim$rpenna/thindisk}.} 
:
flux of radiant energy off the upper surface of the disk, $F$, surface
density, $\Sigma$, disk thickness, $H$, rest mass density in the local
rest frame, $\rho$, temperature, $T$, and radial velocity in the
locally nonrotating frame, $v^{\hat{r}}$. In addition to quantities
that appear explicitly in the equations of structure, we calculate the
optical depth at the center of the disk,
\begin{equation}
\tau=\opacity \Sigma,
\end{equation}
and the characteristic timescale for the gas to move inward from
radius $r$ to the inner edge of the disk,
\begin{equation}
\Delta t(r) = - r/v^{\hat{r}}.
\end{equation}

The disk outside the ISCO can be divided into 4 regions: an ``outer
region'' (large radii) in which gas pressure dominates over radiation
pressure, and in which the opacity is predominantly free-free; a
``middle region'' (smaller radii) in which gas pressure dominates over
radiation pressure, but opacity is predominantly due to electron
scattering; an ``inner region'' (even smaller radii) in which
radiation pressure dominates over gas pressure, and opacity is
predominantly due to electron scattering; and an ``edge region''
(smallest radii) where gas pressure again dominates over radiation
pressure, and opacity is predominantly due to electron scattering.

The NT model does not include the edge region explicitly.  However it
must exist because the no-torque boundary condition implies radiant
flux and radiation pressure go to zero at the ISCO.  The NT inner region
surface density is singular at the ISCO because of
this inconsistency.  The edge region surface density is finite in the
NT limit.

The solutions are functions of the dimensionless radial coordinate
$x=\sqrt{r/M}$.  Calligraphic letters denote functions of $x$ and $a$
with value unity far from the hole.  A subscript $0$ indicates the
quantity is evaluated at the ISCO.

\begin{widetext}  

\subsection{Plunging region}

\newcommand{\vrfirst}{\left[\curly{C}_0^{-1}\curly{G}_0^2\curly{V}-1\right]}
\newcommand{\vrsecond}{\left(\alpha^{1/4}\mstar^{-3/4}\mdotstar^{1/2} \right) 
x_0^{-7/4}
\curly{C}_0^{-5/4}\curly{D}_0^{-1}\curly{G}_0^2\curly{V}}
\newcommand{\vrstar}{v_*}

$p=p^{\rm (gas)}$, $\opacity=\opacity_{es}$.  In this region the
equations of structure (\ref{eq:appmdot})-(\ref{eq:opacity}) yield the
following radial profiles:
{\allowdisplaybreaks
\begin{subequations}
\begin{align}
F &= 
 \left(2\times 10^{18} \text{ erg/cm$^2$ sec}\right)
 \left(\alpha^{4/3}\mstar^{-3}\mdotstar{5/3}\right)
 x^{-26/3}x_0^{4/3}\curly{D}^{-5/6}\curly{K}^{4/3}
 \curly{F}_0^{4/3}\curly{G}_0^{-4/3}\vrstar^{-5/3},\\
\Sigma &=  
 \left( 1 \text{ g/cm$^2$}\right) 
 \left( \mstar^{-1} \mdotstar \right)
 x^{-2}\curly{D}^{-1/2}
 \vrstar^{-1},\\
H &= 
 \left(60 \text{ cm}\right)
 \left(\alpha^{1/6}\mstar^{1/2}\mdotstar^{1/3}\right)
 x^{8/3}x_0^{-5/6}\curly{D}^{-1/6}\curly{K}^{1/6}
 \curly{F}_0^{1/6}\curly{G}_0^{-1/6}\curly{R}_0^{-1/2}\vrstar^{-1/3}\\
\rho &=
 \left(0.01 \text{ g/cm$^3$}\right)
 \left(\alpha^{-1/6}\mstar^{-3/2}\mdotstar^{2/3}\right)
 x^{-14/3}x_0^{5/6}\curly{D}^{-1/3}\curly{K}^{-1/6}
 \curly{F}_0^{-1/6}\curly{G}_0^{1/6}\curly{R}_0^{1/2}\vrstar^{-2/3}\\
T &=
 \left(2\times 10^5 \text{ K}\right)
 \left(\alpha^{1/3}\mstar^{-1}\mdotstar^{2/3}\right)
 x^{-8/3}x_0^{1/3}\curly{D}^{-1/3}\curly{K}^{1/3}
 \curly{F}_0^{1/3}\curly{G}_0^{-1/3}\vrstar^{-2/3}\\
\tau_{es} &=  
 0.3
 \left( \mstar^{-1} \mdotstar \right)
 x^{-2}\curly{D}^{-1/2}
 \vrstar^{-1},\\
\Delta t(r) &= 
 \left(1\times 10^{-5} \text{ sec} \right)
 \left(\mstar \right)x^2 \vrstar^{-1},\\
\vr & = -\left(3\times 10^{10} \text{ cm/sec} \right)\vrstar.
\end{align}
\end{subequations}
}
We have defined the dimensionless radial velocity profile:
\begin{equation}
\vrstar=\left(\vrfirst+\left(7\times 10^{-3}\right)\vrsecond\right)^{1/2}
\end{equation}
The term in square brackets dominates near the horizon.  The term
proportional to $\mdot$ dominates near the ISCO.  At the horizon the
radial velocity is $c$ and at the ISCO it is the sound speed.  In the
limit $\mdot/\mdot_{\rm edd}\rightarrow 0$, we may set
$\vrstar=\vrfirst^{1/2}$ and the gas is released from rest at the
ISCO.

\subsection{Edge region}
\label{sec:edge}

$p=p^{\rm (gas)}$, $\opacity=\opacity_{es}$.  In this region the
equations of structure (\ref{eq:appmdot})-(\ref{eq:opacity}) yield:
{\allowdisplaybreaks
\begin{subequations}
\begin{align}
F &= 
 \left(0.6\times 10^{26} \text{ erg/cm$^2$ sec}\right)
 \left(\mstar^{-2}\mdotstar\right)
 x^{-6}\curly{B}^{-1}\curly{C}^{-1/2}\Phi,\\
\Sigma &=  
 \left( 5\times 10^4 \text{ g/cm$^2$}\right) 
 \left( \alpha^{-4/5}\mstar^{-2/5} \mdotstar^{3/5} \right)
 x^{-6/5}\curly{B}^{-4/5}\curly{C}^{-1/2}\curly{D}^{-4/5}\Phi^{3/5},\\
H &= 
 \left(3\times 10^3 \text{ cm}\right)
 \left(\alpha^{-1/10}\mstar^{7/10}\mdotstar^{1/5}\right)
 x^{21/10}\curly{A}\curly{B}^{-6/5}\curly{C}^{1/2}\curly{D}^{-3/5}\curly{S}^{-1/2}\Phi^{1/5},\\
\rho &=
 \left(10 \text{ g/cm$^3$}\right)
 \left(\alpha^{-7/10}\mstar^{-11/10}\mdotstar^{2/5}\right)
 x^{-33/10}\curly{A}^{-1}\curly{B}^{3/5}\curly{D}^{-1/5}\curly{S}^{1/2}\Phi^{2/5},\\
T &=
 \left(3\times 10^8 \text{ K}\right)
 \left(\alpha^{-1/5}\mstar^{-3/5}\mdotstar^{2/5}\right)
 x^{-9/5}\curly{B}^{-2/5}\curly{D}^{-1/5}\Phi^{2/5},\\
\tau_{es} &=  
 \left(2\times 10^4\right)
 \left(\alpha^{-4/5} \mstar^{-2/5} \mdotstar^{3/5} \right)
 x^{-6/5}\curly{B}^{-3/5}\curly{C}^{1/2}\curly{D}^{-4/5}\Phi^{3/5},\\
\frac{\tau_{ff}}{\tau_{es}}&=
 \left(0.6\times10^{-5}\right)
 \left(\mstar \mdotstar^{-1}\right)
 x^{3}\curly{A}^{-1}\curly{B}^{2}\curly{D}^{1/2}\curly{S}^{1/2}\Phi^{-1},\\
\Delta t(r) &= 
 \left(0.7 \text{ sec} \right)
 \left(\alpha^{-4/5} \mstar^{8/5}\mdotstar^{-2/5} \right)
 x^{14/5} \curly{B}^{-4/5}\curly{C}^{1/2}\curly{D}^{-3/10}\Phi^{3/5},\\
\vr &=
 -\left(6\times 10^{5} \text{ cm/sec} \right)
 \left(\alpha^{4/5} \mstar^{-3/5}\mdotstar^{2/5} \right)
 x^{-4/5} \curly{B}^{4/5}\curly{C}^{-1/2}\curly{D}^{3/10}\Phi^{-3/5}.
\end{align}
\end{subequations}
}
We have defined a new function:
\begin{equation}
\Phi=\curly{Q}+
\left(0.02\right)
\left(\alpha^{9/8}\mstar^{-3/8}\mdotstar^{1/4}\right)
x^{-1}\curly{B}\curly{C}^{-1/2}
\left(x_0^{9/8}\curly{C}_0^{-5/8}
\curly{G}_0\curly{V}_0^{1/2}\right),
\end{equation}
which controls the shape of the radiant flux profile, $F(r)$.  At
large distances, the first term on the RHS is order unity, and the
second term decays as $x^{-1}$.  Near the ISCO, the first term goes to
zero, and the second term is nonzero and proportional to
$\dot{M}^{1/4}$.  So if $\dot{M}/\dot{M}_{\rm edd}$ is small, then the
second term is small everywhere and we may substitute
$\Phi\rightarrow\mathcal{Q}$.  This gives the NT flux.  Quantities
which depend on $\mathcal{S}$ will still differ from NT because our
vertical gravity prescription is different (c.f. \S\ref{sec:vertg}).
We may revert to the incorrect NT vertical gravity with the substitution
$\mathcal{S}\rightarrow \mathcal{E}$.  With these two substitutions
our model becomes the NT solution outside the plunging region.


\subsection{Inner region}

$p=p^{\rm (rad)}$, $\opacity=\opacity_{es}$.  In this region the
equations of structure (\ref{eq:appmdot})-(\ref{eq:opacity}) yield:
{\allowdisplaybreaks
\begin{subequations}
\begin{align}
F &= 
 \left(0.6\times 10^{26} \text{ erg/cm$^2$ sec}\right)
 \left(\mstar^{-2}\mdotstar\right)
 x^{-6}\curly{B}^{-1}\curly{C}^{-1/2}\Phi,\\
\Sigma &=  
 \left( 20 \text{ g/cm$^2$}\right) 
 \left( \alpha^{-1}\mstar \mdotstar^{-1} \right)
 x^{3}\curly{A}^{-2}\curly{B}^{3}\curly{C}^{1/2}\curly{S}\Phi^{-1},\\
H &= 
 \left(1\times 10^5 \text{ cm}\right)
 \left(\mdotstar\right)
 \curly{A}^{2}\curly{B}^{-3}\curly{C}^{1/2}\curly{D}^{-1}\curly{S}^{-1}\Phi,\\
\rho &=
 \left(1\times 10^{-4} \text{ g/cm$^3$}\right)
 \left(\alpha^{-1}\mstar \mdotstar^{-2}\right)
 x^{3}\curly{A}^{-4}\curly{B}^{6}\curly{D}\curly{S}^{2}\Phi^{-2},\\
T &=
 \left(4\times 10^7 \text{ K}\right)
 \left(\alpha^{-1/4}\mstar^{-1/4}\right)
 x^{-3/4}\curly{A}^{-1/2}\curly{B}^{1/2}\curly{S}^{1/4},\\
\tau_{es} &=  
 8
 \left(\alpha^{-1} \mstar \mdotstar^{-1} \right)
 x^{3}\curly{A}^{-2}\curly{B}^{3}\curly{C}^{1/2}\curly{S}\Phi^{-1},\\
\frac{p^{\rm (gas)}}{p^{\rm (rad)}} &=
 \left( 5\times 10^{-5} \right)
 \left(\alpha^{-1/4}\mstar^{7/4}\mdotstar^{-2}\right)
 x^{21/4}\curly{A}^{-5/2}\curly{B}^{9/2}\curly{D}\curly{S}^{5/4}\Phi^{-2},\label{eq:pratio}\\
\Delta t(r) &= 
 \left(2\times 10^{-4} \text{ sec} \right)
 \left(\alpha^{-1} \mstar^{3}\mdotstar^{-2} \right)
 x^{7} \curly{A}^{-2}\curly{B}^{3}\curly{C}^{1/2}\curly{D}^{1/2}\curly{S}\Phi^{-1},\\
\vr & =  -\left(2\times 10^{9} \text{ cm/sec} \right)
 \left(\alpha \mstar^{-2}\mdotstar^{2} \right)
 x^{-5} \curly{A}^{2}\curly{B}^{-3}\curly{C}^{-1/2}\curly{D}^{-1/2}\curly{S}^{-1}\Phi.\\
\end{align}
\end{subequations}
}
The boundaries between the edge, inner, and middle regions can be
computed from the ratio of pressures (\ref{eq:pratio}).

\subsection{Middle region}

$p=p^{\rm (gas)}$, $\opacity=\opacity_{es}$.  The solution is the same
as the edge region solution (c.f. \S\ref{sec:edge}).

\subsection{Outer region}

$p=p^{\rm (gas)}$, $\opacity=\opacity_{ff}$.  In this region the
equations of structure (\ref{eq:appmdot})-(\ref{eq:opacity}) yield:
{\allowdisplaybreaks
\begin{subequations}
\begin{align}
F &= 
 \left(0.6\times 10^{26} \text{ erg/cm$^2$ sec}\right)
 \left(\mstar^{-2}\mdotstar\right)
 x^{-6}\curly{B}^{-1}\curly{C}^{-1/2}\Phi,\\
\Sigma &=  
 \left( 2\times 10^5 \text{ g/cm$^2$}\right) 
 \left( \alpha^{-4/5}\mstar^{-1/2} \mdotstar^{7/10} \right)
 x^{-3/2}\curly{A}^{1/10}\curly{B}^{-4/5}\curly{C}^{1/2}\curly{D}^{-17/20}\curly{S}^{-1/20}\Phi^{7/10},\\
H &= 
 \left(9\times 10^2 \text{ cm}\right)
 \left(\alpha^{-1/10}\mstar^{3/4}\mdotstar^{3/20}\right)
 x^{9/4}\curly{A}^{19/20}\curly{B}^{-11/10}\curly{C}^{1/2}\curly{D}^{-23/40}\curly{S}^{-19/40}
 \Phi^{3/20},\\
\rho &=
 \left(80 \text{ g/cm$^3$}\right)
 \left(\alpha^{-7/10}\mstar^{-5/4}\mdotstar^{11/20}\right)
 x^{-15/4}\curly{A}^{-17/20}\curly{B}^{3/10}\curly{D}^{-11/40}
 \curly{S}^{17/40}\Phi^{11/20},\\
T &=
 \left(8\times 10^7 \text{ K}\right)
 \left(\alpha^{-1/5}\mstar^{-1/2}\mdotstar^{3/10}\right)
 x^{-3/2}\curly{A}^{-1/10}\curly{B}^{-1/5}\curly{D}^{-3/20}\curly{S}^{1/20}\Phi^{3/10},\\
\tau_{ff} &=  
 \left(2\times 10^2\right)
 \left(\alpha^{-4/5} \mdotstar^{1/5} \right)
 \curly{A}^{-2/5}\curly{B}^{1/5}\curly{C}^{1/2}\curly{D}^{-3/5}
 \curly{S}^{1/5}\Phi^{1/5},\\
\frac{\tau_{ff}}{\tau_{es}}&=
 3\times 10^{-3}
 \left(\mstar^{1/2} \mdotstar^{-1/2}\right)
 x^{3/2}\curly{A}^{-1/2}\curly{B}^{2/5}\curly{D}^{1/4}\curly{S}^{1/4}\Phi^{-1/2},\\
\Delta t(r) &= 
 \left(2 \text{ sec} \right)
 \left(\alpha^{-4/5} \mstar^{3/2}\mdotstar^{-3/10} \right)
 x^{5/2} \curly{A}^{1/10}\curly{B}^{-4/5}\curly{C}^{1/2}\curly{D}^{-7/20}
 \curly{S}^{-1/20}\Phi^{7/10},\\
\vr & = -\left(2\times 10^{5} \text{ cm/sec} \right)
 \left(\alpha^{4/5} \mstar^{-1/2}\mdotstar^{3/10} \right)
 x^{-1/2} \curly{A}^{-1/10}\curly{B}^{4/5}\curly{C}^{-1/2}\curly{D}^{7/20}
 \curly{S}^{1/20}\Phi^{-7/10}.\\
\end{align}
\end{subequations}
}
\end{widetext}

\section{Comparison with slim disk models}
\label{sec:slim}

Slim disk solutions include advection.
They are computed numerically subject
to the condition that the flow pass smoothly through a sonic point.
The position of the sonic point is free to vary.  A model for energy
transport in the vertical direction, including radiative transport and
convection, is coupled to the radial equations.  For a complete
description see \cite{sadowski11}.

\begin{figure}
\includegraphics[width=3.3in,clip]{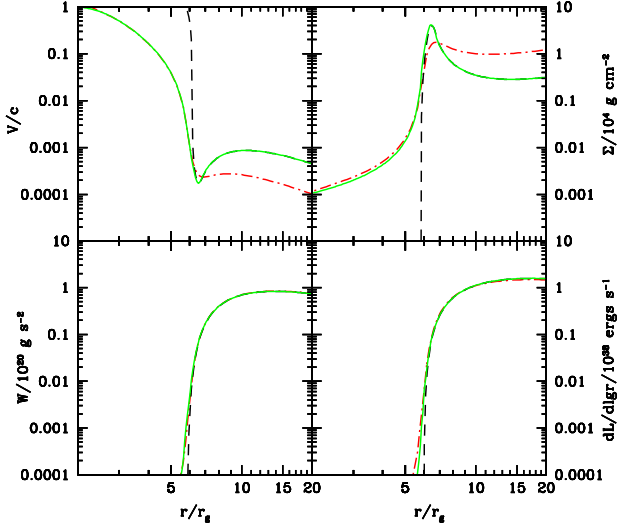}
\caption{Radial velocity, surface density, stress, and luminosity
  versus radius for three disk models: NT (dashed black), slim disk
  (dot-dashed red), and our generalized thin disk (solid green).
  These solutions have $M=10 M_\odot$, $a_*=0$, $\alpha=0.1$, and
  $\dot{M}/\dot{M}_{\rm edd}=0.3$.  The NT solution
  terminates at the ISCO but the slim disk and our model continue to
  the event horizon.}
\label{fig:slima0}
\end{figure}

\begin{figure}
\includegraphics[width=3.3in,clip]{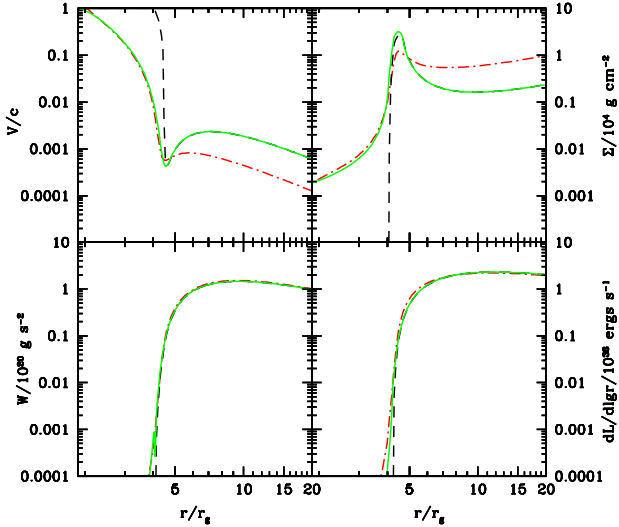}
\caption{Same as Figure \ref{fig:slima0} but with $a_*=0.5$.}
\label{fig:slima6}
\end{figure}

When in the thin disk regime, $h \ll \alpha$ and
$\dot{M}/\dot{M}_{\rm edd}<0.3$, the sonic point is near the ISCO and
advection can be neglected (c.f. \S\ref{sec:bc}).  This
enables the analytical solution of \S\ref{sec:gtd}.  So we expect our model
and slim disk models to be similar under these conditions.  We make
this comparison in Figures \ref{fig:slima0} and \ref{fig:slima6}.  The
NT disk is also included.

The disk solutions in Figure \ref{fig:slima0} have 
$M=10 M_\odot$, $a_*=0$, $\alpha=0.1$ and $\dot{M}/\dot{M}_{\rm
  edd}=0.3$.  Our model and slim disk solutions are in good agreement
inside the ISCO where the NT model does not extend.  The slim disk
surface density is larger outside the ISCO because it includes both
radiation and gas pressure (which are comparable there), while the
analytical disk models include only radiation pressure at these
radii. All three models eventually converge at large radii. Figure
\ref{fig:slima6} is the same except with $a_*=0.5$.

\section{Comparison with GRMHD simulations}
\label{sec:grmhd}

In this section, we compare the fiducial, $a_*=0$ GRMHD simulation of
\citet{kulkarni11} with analytical thin disk solutions.  The target
entropy in the GRMHD cooling function was $K_c=0.00034$ and the
gas-to-magnetic pressure ratio of the initial fields in the simulation
was $\beta_i=100$.  These parameters play similar roles to $\dot{M}$
and $\alpha$ in the thin disk solutions.  The simulation reached
$t=26300M$.  At large radii, the viscous timescale is long and the
solution is not converged.  The estimates in \citet{rfp10} suggest
steady state is reached out to $r= 9M$.

\subsection{Hydrostatic equilibrium of GRMHD disks}

In hydrostatic equilibrium, opening angle is
related to pressure and density by \eqref{eq:hydroeq}:
\begin{equation}\label{eq:hydrohr}
h=\sqrt{\frac{p}{\rho}}\frac{r}{\abr},
\end{equation}
where $\abr^2 =u_\phi^2-a^2\left(u_t-1\right)$ \citep{abr97}.  Our
thin disk solutions assume hydrostatic equilibrium even inside the
plunging region, so we first check whether this is a good description
of GRMHD disks.

\begin{figure}
\includegraphics[width=3.0in,clip]{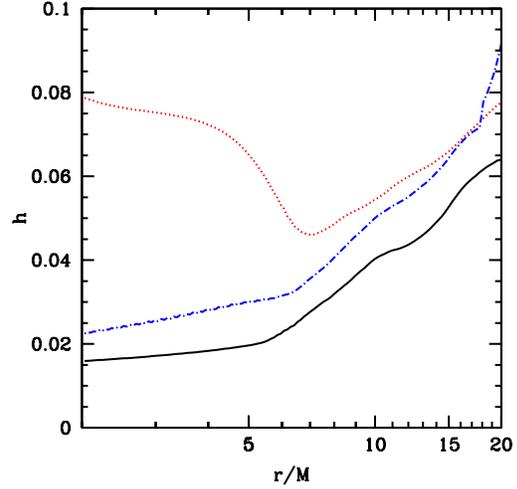}
\caption{Unnormalized, $\rho$-weighted opening angle $h^{(\rm rms)}$
  (dotted red), normalized, $\rho^2$-weighted opening angle $h^{(\rm
  rms)}_2$ (solid black), and opening angle expected from hydrostatic
  equilibrium (dot-dashed blue).   The opening angle expected from
  hydrostatic equilibrium is within $30\%$ of $h^{(\rm rms)}_2$ at all
  converged radii.}
\label{fig:grmhdhr}
\end{figure}

A popular definition of opening angle for GRMHD disks is
\citep{rfp10,nkh10,shafee08}:
\begin{equation}\label{eq:unthetarms}
h^{(\rm rms)} \equiv 
\left(
\frac{\int \left( \theta-\pi/2 \right)^2 \rho \sqrt{-g}  dt d\theta d\phi    }
{\int \rho \sqrt{-g}   dt d\theta d\phi    }
\right)^{1/2}.
\end{equation}
In Figure \ref{fig:grmhdhr}, we compare $h^{(\rm rms)}$ for the
fiducial GRMHD disk (dotted red) with the RHS of \eqref{eq:hydrohr},
the opening angle expected from hydrostatic equilibrium (dot-dashed
blue).  The later is computed from the rest mass density and total
pressure at the midplane. (Replacing midplane values with
density-weighted vertical averages has little effect.)  Hydrostatic
equilibrium appears to be a bad approximation in the plunging region,
where it gives the wrong opening angle by as much as a factor of 4.

The simulation's high density, gas pressure dominated disk is
surrounded by a low density, magnetically supported corona.  These two
regions have different scale heights and the apparent deviations from
hydrostatic equilibrium could be a result of mixing them in the
definition of opening angle.  Our thin disk model only applies to the
disk region, so we would like to minimize the contribution of the
corona.  We can do this by weighting the integrals in
\eqref{eq:unthetarms} with higher powers of $\rho$, because this
concentrates attention on high density regions.  Unfortunately, the
opening angle then depends on this choice: higher powers of $\rho$
give smaller opening angles.  To get an invariant measure, we
normalize the opening angle as follows.  The vertical density profile
of a polytropic gas in hydrostatic equilibrium is
\begin{equation}\label{eq:vertrho}
\rho(z)=\rho(z=0)\left(1-\frac{\left(z/r\right)^2}{\mathcal{H}^2}\right)^N,
\end{equation}
where $\mathcal{H}$ is the opening angle.  The simulation has
$\Gamma=1+1/N=4/3$, so $N=3$.  We normalize our definition of opening
angle such that it returns $\mathcal{H}$ when given the analytical
solution \eqref{eq:vertrho}.  So the normalized, $\rho^2$-weighted
opening angle is
\begin{equation}\label{eq:thetarms}
h^{(\rm rms)}_2 \equiv \frac{\sqrt{15}}{3}
\left(
\frac{\int \left( \theta-\pi/2 \right)^2 \rho^2 \sqrt{-g}  dt d\theta d\phi    }
{\int \rho^2 \sqrt{-g}   dt d\theta d\phi    }
\right)^{1/2},
\end{equation}
where $\sqrt{15}/3$ is the the normalization defined by
\eqref{eq:vertrho}.  

We plot $h^{(\rm rms)}_2$ for the fiducial GRMHD simulation in Figure
\ref{fig:grmhdhr} (solid black).  It is within $30\%$ of the opening
angle expected from hydrostatic equilibrium (dot-dashed blue) at all
converged radii.  Emphasizing the disk over the corona has removed the
discrepancy and shows that hydrostatic equilibrium is a good approximation in
the disk region.  From now on, we define the GRMHD opening angle to be
$h=h^{(\rm rms)}_2$.


\subsection{The boundary between disk and corona}
\label{sec:diskvcorona}

\begin{figure}
\includegraphics[width=3.3in,clip]{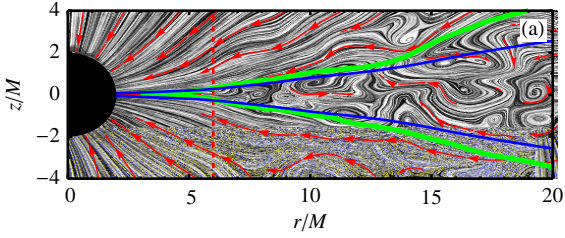}
\caption{A snapshot of the velocity streamlines in the saturated state
of the fiducial GRMHD disk simulation.  The $\beta=1$ contour (green)
divides the high density, weakly magnetized disk from the low density,
highly magnetized coronal and plunging regions.  Magnetic tension
stabilizes strong fields against the MRI, so the corona is laminar and
the disk is turbulent.  The contour $2h$ (blue) roughly
corresponds to $\beta=1$.  The dashed red line marks the ISCO.}
\label{fig:grmhdstream}
\end{figure}

The boundary between disk and corona can be identified with the
contour where the gas-to-magnetic pressure ratio $\beta=1$, where the
pressure switches from predominantly gas to magnetic.  This is plotted
in green in Figure \ref{fig:grmhdstream}.  The corona is located
approximately $2h$ away from the midplane (blue contour).  The flow is
turbulent in the disk and laminar in the corona because strong fields
are stabilized against the MRI by magnetic tension \citep{pessah05}.
Working in the Boussinesq approximation and ignoring magnetic
curvature terms, \citet{bal91,bh98} argued the MRI cannot operate when
$\beta<1$.

The plunging region inside the ISCO is similar to the corona, although
it is highly magnetized for a different reason.  The corona is highly
magnetized because magnetic buoyancy raises fields out of the disk.
The region inside the ISCO is highly magnetized because plunging fluid
stretches frozen-in field lines.

\subsection{Effective $\alpha$}

Viscosity in the simulation is generated by MRI-driven turbulence.  We
define the effective $\alpha$:
\begin{equation}
\alpha \equiv \frac{\left<W\right>}{2 \left<p\right> hr}.
\end{equation}
The height integrated stress, $\left< W\right>$, is computed from the GRMHD
stress-energy tensor \citep{rfp10}:
\begin{equation}
\left< W \right>
=\frac{1}{2\pi \Delta t}
\int_{t}^{t+\Delta t} 
\int_{\pi/2+2h}^{\pi/2-2h} 
\int_0^{2\pi} 
T_{\hat{r}\hat{\phi}}^{\rm GRMHD} 
\sqrt{-g} dt d\theta d\phi.
\end{equation}
The pressure, $\left< p \right>$, is a $\rho^2$-weighted height average:
\begin{equation}
\left< p \right>=
\frac{
\int
p \rho^2 \sqrt{-g} dt d\theta d\phi    }
{
\int
\rho^2 \sqrt{-g} dt d\theta d\phi    }
.
\end{equation}

\begin{figure}
\includegraphics[width=3.3in,clip]{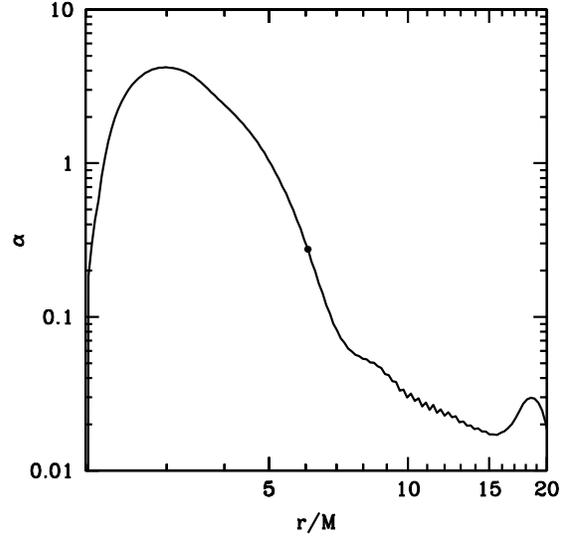}
\caption{Effective $\alpha$ of the fiducial GRMHD disk simulation.
  The ISCO is marked with a black dot.  There is a sharp rise inside
  the ISCO, where the plunging fluid stretches the field lines.
  Unlike slim disk and thin disk models, the GRMHD disk does not have a constant $\alpha$.
}
\label{fig:grmhdalpha}
\end{figure}

The effective $\alpha$ is plotted as a function of radius in Figure
\ref{fig:grmhdalpha}.  We use $\alpha=0.3$, the effective $\alpha$ at
the ISCO, to compare thin disk solutions with the GRMHD disk.

\subsection{Comparison}
\label{sec:comparison}

\begin{figure}
\includegraphics[width=3.3in,clip]{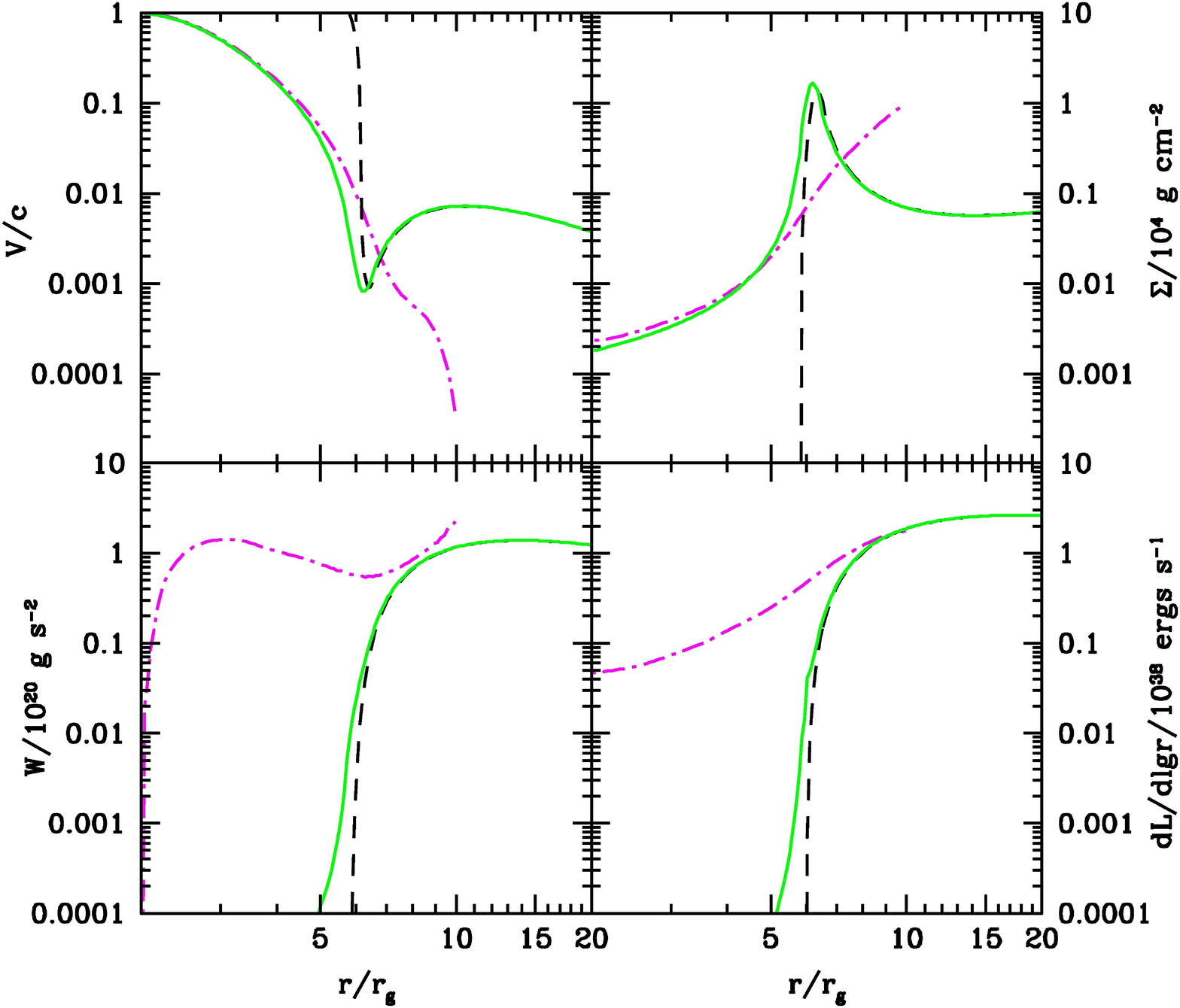}
\caption{Radial velocity, surface density, stress, and luminosity
  versus radius for the fiducial GRMHD disk simulation (dot-dashed
  red) and NT (dashed black) and generalized thin disk models (solid
  green) with $\alpha=0.3$ and $\dot{M}/\dot{M}_{\rm edd}=0.5$.  The
  GRMHD curves are truncated at $r=9M$, beyond which the simulations
  have not reached steady state.
  \citep{kulkarni11}. }
\label{fig:grmhdGTD}
\end{figure}

Figure \ref{fig:grmhdGTD} compares the GRMHD disk to thin disk
solutions with $M=10M_\odot$.  We assume $\dot{M}/\dot{M}_{\rm
edd}=0.5$, which \citet{kulkarni11} estimated to be the effective
accretion rate of this simulation.  The GRMHD profiles are only shown
out to $r=9M$ because the simulations are not converged beyond this.
Simulation data is time-averaged over the steady state period from
$t=21000M-26300M$ \citep{kulkarni11,rfp10}.

The radial velocity and surface density are well-described in the
plunging region.  This validates the thin disk approximations
made in \S\ref{sec:approx}.  In particular, the GRMHD plunging region is well
approximated by hydrostatic equilibrium, with inner edge and
sonic point at the ISCO, and geodesic motion.

Outside the ISCO the stress is primarily turbulent in origin.  Inside
the ISCO, it is generated by mean magnetic fields, which are stretched
and amplified by the plunging fluid.  Magnetic field reconnection at
the grid scale and shocks \citep{beskin05} create luminosity inside
the ISCO.  The analytical disk models do not contain magnetic fields,
so this physics is not captured.  This is why the GRMHD stress and
luminosity are not well described by the analytical models inside the
ISCO.  GRMHD disks are thicker inside the ISCO for the same reasons.

\section{Conclusions}
\label{sec:conclude}

We have developed an analytical model for thin disk accretion in the
Kerr metric which generalizes the NT model in three ways: (i) it
incorporates nonzero stresses at the inner edge of the disk, (ii) it
extends into the plunging region, and (iii) it uses the correct
vertical gravity formula.  The free parameters are unchanged.  Our
model is a special case of slim disk solutions, in the regime $h \ll
\alpha$ and $\dot{M}_{\rm edd}/\dot{M}<0.3$.  Under these conditions,
energy advection is less important than the stress at the inner edge of the
disk, and the inner edge, sonic point, and ISCO are at approximately
the same position.

The boundary condition is supplied by setting
the radial velocity at the ISCO equal to the sound speed.  In the
limit $h \rightarrow 0$, this reduces to the NT zero-stress boundary
condition.  Outside the ISCO, the stress and radiant flux are the sum
of the NT prediction and a correction term which incorporates the
stress at the ISCO.  Inside the ISCO, fluid plunges into the black
hole and the motion is approximately geodesic.  This enables us to
estimate the pressure, and then the stress and radiant flux, of the
plunging gas.  Throughout we assume the fluid is in vertical
hydrostatic equilibrium and the stress is described by an
$\alpha$-viscosity.  The model compares favorably with slim disk
solutions.

We fit our disk solutions to a GRMHD disk simulation.  
We argued that the $\beta=1$ contour is a natural boundary between the
disk and coronal regions in GRMHD simulations.  Fluid in the disk is
turbulent.  Outside the disk, where the field is strong and the MRI
cannot operate, the flow is laminar.

The velocity and surface density are well-modeled inside the ISCO.
This validates our assumptions that the fluid is in hydrostatic
equilibrium and the velocities are nearly geodesic.

The GRMHD plunging region stress is larger than the stress in
hydrodynamic models.  The stresses are carried by large scale, mean
magnetic fields.  Some of this stress is dissipated by magnetic
reconnection at the grid scale.  The slim and thin disk models do not
include magnetic fields, so they cannot model this.

Black hole spin parameters can be measured by modeling X-ray
spectra using the NT accretion disk
\citep{zcc97,shafee06,ddb06,gmlnsrodes09,gmsncbo10,2009ApJ...701L..83S}.
The NT model assumes advection is negligible, the disk inner edge is
at the ISCO, and there are no torques at the ISCO.  Slim disk
solutions are available which do not make these assumptions.
\citet{straub11} replaced the NT disk with a slim disk model and
revisited the spin estimate of LMC X-3.  They were unable to improve
the estimate because theoretical improvements in the slim disk model
were smaller than observational errors.

Observational errors in black hole spin measurements come from
uncertainty in the black hole mass, distance, and disk inclination
\citep{mcc06}.  Current observational uncertainties in spin estimates
are at best $\Delta \astar \pm 0.2$ at low spins and $\Delta \astar
\pm 0.05$ at high spins \citep{gmlnsrodes09}.  These estimates are
made using data with $\dot{M}/\dot{M}_{\rm edd} < 0.3$.
\citet{kulkarni11} created mock data from a GRMHD simulation and
fitted it with a NT disk, and computed an estimate of the spin error
coming from disk theory.  At $\dot{M}/\dot{M}_{\rm edd} \sim 0.5$ and
$\alpha \sim 0.3$, they found spin errors of $\Delta \astar \pm 0.2$ at low
spins and $\Delta \astar \pm 0.01$ at high spins.  Errors increase
with luminosity, so the theoretical uncertainties are always smaller
than the observational ones.

This means the NT disk is sufficient for spin measurements at present.
However, more sophisticated disk models will be needed as black hole
mass, distance, and disk inclination measurements improve.  In the
hierarchy of disk models, the model in this paper contains more
physics than NT but less than a slim disk.  Our model and the slim
disk are similar when $\dot{M}/\dot{M}_{\rm edd}>0.3$ and $h<\alpha$,
but our model is analytical, so it might be simpler to use in some
cases.

\citet{mcc06} introduced the selection criterion $\dot{M}/\dot{M}_{\rm
edd} < 0.3$ when they measured the spin of the black hole GRS
1915+105.  Black hole X-ray binaries have variable luminosities and
the NT model is only valid at low luminosities.  Of the 22
observations of GRS 1915+105 available to \citet{mcc06}, five
satisfied the $\dot{M}/\dot{M}_{\rm
edd} < 0.3$ criterion.  These five observations gave a nearly
consistent spin parameter $\astar > 0.98$.  Observations with
$\dot{M}/\dot{M}_{\rm edd}>0.3$ give inconsistent spin results, but
the NT model is not valid in this regime.

\section*{Acknowledgments}

We thank Stephen Balbus, Ramesh Narayan, and Alexander Tchekhovskoy
for discussions.  This work was supported by an NSF Graduate Research
Fellowship (RFP), NSF grant AST-1041590, NASA grant NNX11AE16G, and by
the NSF through TeraGrid resources provided by NCSA (Abe), LONI
(QueenBee), and NICS (Kraken) under grant numbers TG-AST080025N and
TG-AST080026N.

\appendix

\section{The Kerr metric}
\label{app:metric}

We assume spinning black holes can be described by the Kerr metric and
the accretion disk lies in the equatorial plane of the metric.  The
Kerr metric in Boyer-Lindquist coordinates $(t,r,\theta,\phi)$ in and
near the equatorial plane $(|\theta-\pi/2|\ll 1)$ is:
\begin{align}
ds^2 &=
-\frac{r^2 \Delta}{A}dt^2
+\frac{A}{r^2}\left(d\phi-\omega dt\right)^2 
+\frac{r^2}{\Delta}dr^2+dz^2,\\
\Delta &= r^2-2Mr+a^2,\notag\\
%
A &= r^4 + r^2a^2+2Mra^2,\notag\\
%
\omega &= 2Mar/A.\notag
\end{align}
Here $M$ and $a$ are the mass and specific angular momentum of the
hole.  We have replaced the usual angular coordinate by $z=r\cos\theta\simeq
r\left(\theta-\pi/2\right)$. Define the auxiliary parameters:
\begin{subequations}
\begin{align}
\astar &= a/M \quad\text{(note: $-1\leq a_* \leq +1$)},\\
x_1 &= 2 \cos\left(\cos^{-1}\left({\astar}\right)/3-\pi/3\right),\\
x_2 &= 2 \cos\left(\cos^{-1}\left({\astar}\right)/3+\pi/3\right),\\
x_3 &= -2 \cos\left(\cos^{-1}\left(\astar\right)/3\right),
\end{align}
\end{subequations}
and a dimensionless radial coordinate
\begin{equation}
x   = (r/M)^{1/2}.
\end{equation}
For simplicity in splitting formulae into Newtonian limits plus
relativistic corrections, we shall introduce the following functions
of $x$ and $a_*$ with value unity far from the hole:

{\allowdisplaybreaks
\begin{subequations}
\begin{align}
\curly{A} &= 1+ \astar^2 x^{-4} + 2\astar^2 x^{-6}, \\
\curly{B} &= 1 + \astar x^{-3}, \\
\curly{C} &= 1 - 3 x^{-2} + 2 \astar^2  x^{-3}, \\
\curly{D} &= 1 - 2 x^{-2} + \astar^2  x^{-4}, \\
\curly{E} &= 1+ 4\astar^2 x^{-4} - 4\astar^2  x^{-6}+3\astar^4 x^{-8}, \\
\curly{F} &= 1 -2 \astar x^{-3} + \astar^2  x^{-4}, \\
\curly{G} &= 1 -2 x^{-2} + \astar x^{-3},\\
\curly{H} &= 1-2x^{-2}+2\astar x^{-2} x_0^{-1} \curly{F}_0^{-1} \curly{G}_0,\\
\curly{I} &= \curly{A}-2\astar x^{-6} x_0 \curly{F}_0 \curly{G}_0^{-1},\\
 \curly{J} &= \curly{O}-x^{-2}\curly{I}^{-1}
 \biggl[1-\astar x_0^{-1}\curly{F}_0^{-1}\curly{G}_0
 +\astar^2 x^{-2}\curly{H} \curly{I}^{-1}\notag\\
 & \times \left(1+3x^{-2}-3\astar^{-1}x^{-2}x_0\curly{F}_0\curly{G}_0^{-1}\right)
 \biggr], \\
 \curly{K} &= \left|\curly{A}\curly{J}\left(
 1-x^{-4}\curly{A}^2\curly{D}^{-1}
 \left(
 x_0\curly{F}_0\curly{G}_0^{-1}\curly{O}-2\astar x^{-2}\curly{A}^{-1}
 \right)^2
 \right)^{-1}\right|,\\
\curly{O} &= \curly{H}\curly{I}^{-1},\\
\curly{Q} &= 
  \curly{B}\curly{C}^{-1/2}
  \frac{1}{x}
  \biggl[x-x_0-\frac{3}{2}\astar \ln\left(\frac{x}{x_0}\right)\notag\\
  & -\frac{3\left(x_1-\astar\right)^2}
  {x_1\left(x_1-x_2\right)\left(x_1-x_3\right)}
  \ln\left(\frac{x-x_1}{x_0-x_1}\right)\notag\\
  & -\frac{3\left(x_2-\astar\right)^2}
  {x_2\left(x_2-x_1\right)\left(x_2-x_3\right)}
  \ln\left(\frac{x-x_2}{x_0-x_2}\right)\notag\\
  & -\frac{3\left(x_3-\astar\right)^2}
  {x_3\left(x_3-x_1\right)\left(x_3-x_2\right)}
  \ln\left(\frac{x-x_3}{x_0-x_3}\right)\biggr],\\
\curly{R} &=
 \curly{F}^2\curly{C}^{-1}-\astar^2 x^{-2}
 \left(\curly{G}\curly{C}^{-1/2}-1\right),\\
\curly{S} &=
 \curly{A}^2\curly{B}^{-2}\curly{C}\curly{D}^{-1}\curly{R},\\
%
%
\curly{V} &=  \curly{D}^{-1} 
\biggl[ 1
  +x^{-4}\left(\astar^2-x_0^2 \curly{F}_0^2 \curly{G}_0^{-2}\right)
  +2x^{-6}\left(\astar-x_0\curly{F}_0\curly{G}_0^{-1}\right)
\biggr].
\end{align}
\end{subequations}
}
A subscript $0$ indicates the quantity is evaluated at the innermost
stable circular orbit (ISCO).  Functions $\curly{A}$-$\curly{G}$ and
$\curly{Q}$ are taken from \citet{nt73} and \citet{pt74}.

\subsection{Geodesics}

The non-zero components of the four-velocity, $u^\mu$, for general
equatorial, timelike geodesic motion in the Kerr metric are
\citep{chandraBH}:
\begin{subequations}
\begin{align}
u^t &= 
\frac{1}{\Delta}\left[\left(r^2+a^2+\frac{2a^2M}{r} \right)E 
-\frac{2aM}{r}L \right],  \label{eq:tdot} \\
u^r &= 
\frac{1}{r^2}\left[ r^2 E^2 +\frac{2M}{r}(a E-L)^2
+(a^2E^2-L^2)-\Delta\right], \label{eq:rdot}\\
u^\phi &= 
\frac{1}{\Delta}\left[\left(1-\frac{2M}{r}\right)L
+\frac{2aM}{r}E\right],\label{eq:phidot}
\end{align}
\end{subequations}
where $E$ and $L$ are the conserved specific energy and angular
momentum of the motion.  Circular geodesics have energy per unit mass
\begin{equation}
E=|u_t|=\mathcal{G}/\mathcal{C}^{1/2},\label{eq:Ecirc}
\end{equation}
angular momentum per unit mass
\begin{equation}
L=u_\phi=M^{1/2}r^{1/2}\curly{F}/\curly{C}^{1/2},\label{eq:Lcirc}
\end{equation}
and angular velocity
\begin{equation}
\Omega 
= \frac{u^\phi}{u^t} 
=\frac{M^{1/2}}{r^{3/2}+aM^{1/2}}
=\frac{M^{1/2}}{r^{3/2}}\frac{1}{\curly{B}}.\label{eq:omega}
\end{equation}
Circular geodesics are unstable inside the ISCO.  The radius of the ISCO
is:
\begin{align}
r_0/M &= 
3+Z_2-
\left[\left(3-Z_1\right)\left(3+Z_1+2Z_2\right)\right]^{1/2},\\
Z_1   &=
1+\left(1-\astar^2\right)^{1/3}
\left[\left(1+\astar\right)^{1/3}+\left(1-\astar\right)^{1/3}\right],\notag\\
Z_2   &=
\left(3\astar^2+Z_1^2\right)^{1/2}\notag
\end{align}
The linear velocity of a circular orbit relative to a locally nonrotating
observer \citep{bardeen72} is
\begin{equation}
V_{(\phi)}=\frac{A}{r^2\Delta^{1/2}}\left(\Omega-\omega\right).
\end{equation}
The Lorentz factor corresponding to this linear velocity is
\begin{equation}
\gamma=\left(1-V_{(\phi)}^2\right)^{-1/2}.
\end{equation}
The only nonzero components of the fluid frame shear tensor for the
congruence of circular, equatorial geodesics are
\begin{equation}
\sigma_{\hat{r}\hat{\phi}}
=\sigma_{\hat{\phi}\hat{r}}
=\frac{1}{2}\frac{A}{r^3}\gamma^2\Omega_{,r}.\label{eq:shear}
\end{equation}

\section{Disk structure equations}
\label{app:disk}

\subsection{Definitions}

The stress-energy tensor of a relativistic fluid is
\begin{equation}
\mathbf{T}=\rho(1+\Pi)\mathbf{u}\otimes\mathbf{u}+\mathbf{t}+
           \mathbf{u}\otimes\mathbf{q}+\mathbf{q}\otimes\mathbf{u},
\end{equation}
where $\rho$ is rest mass density in the local rest frame of the
baryons (LRF), $\Pi$ is the specific internal energy in the LRF,
$\mathbf{t}$ is the stress tensor in the LRF, and $\mathbf{q}$ is the
energy flux relative to the LRF.  We make the thin disk
approximation $\Pi=0$ (c.f. \S\ref{sec:bc}).

The disk structure equations are expressed in terms of the surface
density of the disk,
\begin{equation}
\Sigma=\int^{+H}_{-H} \rho dz = 2 \rho H,
\end{equation}
the integrated shear stress,
\begin{equation}
W=\int^{+H}_{-H} t_{\hat{\phi}\hat{r}} dz = 2 t_{\hat{\phi}\hat{r}} H,
\end{equation}
the radial velocity of the gas in the locally non-rotating frame,
\begin{equation}
v^{\hat{r}}=\sqrt{g_{rr}}u^r,
\end{equation}
and the the flux of radiant energy off the upper face of the disk,
\begin{equation}
F=q^{\hat{z}}(z=+H)=q^{\hat{z}}(z=-H),
\end{equation}
where the disk scale height, $H$, is defined by $h=H/r$.
\subsection{Radial structure equations}

The radial structure of the disk is controlled by conservation of
baryon number, conservation of angular momentum, and conservation of
energy:
\begin{align}
\left(\rho u^\mu\right)_{;\mu}  &= 0,\label{eq:continuity} \\
\left(T^\mu_\phi \right)_{;\mu} &= 0,\label{eq:angularmomentum} \\
\left(T^\mu_t\right)_{;\mu}     &= 0.\label{eq:energycons}
\end{align} 
Integrating (\ref{eq:continuity}) gives the accretion rate of a
stationary disk:
\begin{equation}
\dot{M}=-2\pi r \Sigma u^r=(\text{constant independent of $r$ and
  $t$}).
\label{eq:appmdot}
\end{equation}
Combining angular momentum and energy conservation gives
(c.f. \S\ref{sec:bc}):
\begin{equation}
F=-\shear W.\label{eq:fluxstress}
\end{equation}
Combining all three conservation laws gives an integral solution for the flux:
\begin{equation}\label{eq:PTintegral}
4\pi r\frac{(E-\Omega L)^2}{-\Omega_{,r}}F/\dot{M}=
\int_{r_{\rm 0}}^r (E-\Omega L)L_{,r}dr 
+ C.
\end{equation}
\citet{pt74} give an analytical formula for the integral on the RHS
when $r \geq r_{\rm 0}$.  The integration constant $C$ is related to
the flux at the ISCO. The NT no-torque boundary condition is $C=0$.
We allow nonzero $C$.

\subsection{Vertical structure equations}

The vertical structure of the disk is controlled by pressure balance
(c.f. \S\ref{sec:vertg}),
\begin{align}\label{eq:hydroeq}
-\frac{p}{\rho}+h^2\frac{\abr^2}{r^2} = 0, \\
\abr^2 =u_\phi^2-a^2\left(u_t-1\right),\notag
\end{align}
the Shakura-Sunyaev $\alpha$-viscosity prescription,
\begin{equation}\label{eq:alphastress}
t_{\hat{\phi} \hat{r}}=\alpha p
\end{equation}
radiative energy transport,
\begin{equation}
bT^4=\opacity \Sigma F,
\end{equation}
the equation of state,
\begin{align}
p &= p^{\rm (rad)}+p^{\rm (gas)}, \\ 
p^{\rm (rad)} &= \frac{1}{3}bT^4,\notag\\
p^{\rm (gas)} &= \rho\left(T/m_p\right),\notag
\end{align}
and the opacity law,
\begin{align}
\opacity       &= \opacity_{ff}+\opacity_{es},\label{eq:opacity}\\
\opacity_{ff}  &=\left(0.64\times 10^{23}\right)
\left(\frac{\rho}{\text{g/cm$^3$}}\right)
\left(\frac{T}{K}\right)^{-7/2}\frac{\text{cm$^2$}}{g},\notag\\
\opacity_{es} &=0.40 \frac{\text{cm$^2$}}{\text{g}}\notag.
\end{align}

\subsection{Solving for the disk structure}
\label{sec:appsolution}

At this point the disk structure is defined by seven equations
(\ref{eq:appmdot})-(\ref{eq:opacity}) for nine unknowns $E,
L, u^r, W, F, \Sigma, p, h, T, \opacity$ and four free parameters $M,
a, \dot{M}, \alpha$.  The integration constant $C$ is fixed by the
boundary condition (\ref{eq:intconst}). 

To close the problem we need two more relations.  These are
prescriptions for $E$ and $L$.  Outside the ISCO, the disk nearly
follows circular geodesics so $E$ and $L$ are (\ref{eq:Ecirc}) and
(\ref{eq:Lcirc}).  Inside the ISCO, the disk follows non-circular
plunge trajectories with constant $E$ and $L$ defined in
\S\ref{sec:plunge}.

The fluid flow is slightly non-geodesic because it is acted upon by
stresses.  These are small deviations because the disk is thin and
there are several ways of treating them.  The different prescriptions
are equivalent in the thin disk limit, so we choose the simplest.

Outside the ISCO, we use the angular velocity of circular geodesics
(\ref{eq:omega}), but do not enforce $u^r=0$ (as would be required if
the flow were truly geodesic by (\ref{eq:rdot})).  We use this angular
velocity when computing the shear tensor (\ref{eq:shear}).

Inside the ISCO, we use the geodesic velocities
(\ref{eq:tdot})-(\ref{eq:phidot}), but do not assume the radiant flux
integral (\ref{eq:PTintegral}) is zero (as would be required if $L$ were truly
constant). This eliminates one independent variable from the problem
in the plunging region (because $u^r$ is fixed by the geodesic
equations) and one of the disk structure equations (because we do not
enforce (\ref{eq:PTintegral})).  

Throughout most of the plunging region, the angular velocity exceeds
the radial velocity, so we use (\ref{eq:phidot}) and (\ref{eq:shear})
to compute the shear.  This fails near the photon orbit, but the thin
disk approximations are expected to break down there (\S\ref{sec:bc}).

Explicit solutions are in \S\ref{sec:gtd}.

\section{Scaling of compression, advection, and boundary stress terms with
$\alpha$ and \lowercase{$h$}}
\label{app:bc}

The law of energy conservation (\ref{eq:energycons}) can be rewritten
\citep{nt73}:
\begin{equation}\label{eq:energy}
\rho \frac{d\Pi}{d\tau} +\nabla\cdot \mathbf{q} =
-\sigma_{\alpha\beta}t^{\alpha\beta}
-\frac{1}{3}\theta t^\alpha_\alpha-\mathbf{a}\cdot \mathbf{q}.
\end{equation}
We have introduced the convective derivative
$d/d\tau\equiv\mathbf{u}\cdot\nabla$, the scalar expansion
$\theta\equiv\nabla\cdot\mathbf{u}$, the acceleration vector
$\mathbf{a}\equiv \nabla_{\mathbf{u}}\mathbf{u}$, the shear tensor
\begin{equation}
\sigma_{\alpha\beta}\equiv
\frac{1}{2}\left(
u_{\alpha;\mu}h^\mu_\beta
+u_{\beta;\mu}h^\mu_\alpha\right)
-\frac{1}{3}\theta h_{\alpha\beta},
\end{equation} 
and the projection tensor
\begin{equation}
h_{\alpha\beta}\equiv g_{\alpha\beta}+u_\alpha u_\beta.
\end{equation}
Each term in (\ref{eq:energy}) has a simple physical interpretation.
The term $\mathbf{a}\cdot\mathbf{q}$ is a special relativistic
correction associated with the inertia of the flowing energy
$\mathbf{q}$.  We assume $\mathbf{q}$ is directed entirely along $z$
and $u^z=0$, so $\mathbf{a}\cdot\mathbf{q}=0$.  

The remaining terms on the RHS of (\ref{eq:energy}) correspond to
energy generation by shear stresses,
$-\sigma_{\alpha\beta}t^{\alpha\beta}$, and by compression, $-1/3
\theta t^\alpha_\alpha$.  The sink terms on the LHS of
(\ref{eq:energy}) describe energy advection, $-\rho d\Pi /d\tau$,
and radiative losses, $\nabla \cdot \mathbf{q}$.

After height integrating and normalizing by $\dot{M}$, the compression
term scales as
\begin{equation}\label{eq:compress}
\frac{\theta t^\alpha_\alpha h}{\dot{M}}
\propto \frac{u^r_{,r}p h}{\rho u^rh} 
\propto h^2.
\end{equation}
Here $\propto$ means proportionality with respect to $h$ and $\alpha$,
which are considered small.  So, for example, $u^r_{,r}/u^r\propto
(u^r/r)/u^r \propto 1$.  In the first step of (\ref{eq:compress}), we
inserted the accretion rate (\ref{eq:appmdot}).  In the second step,
we used the pressure balance relation (\ref{eq:hydroeq}).

The height integrated advection term scales as
\begin{equation}\label{eq:advection}
\frac{\rho d\Pi / d\tau h}{\dot{M}}
\propto \frac{\rho u^r \Pi_{,r} h}{\rho u^r h}
\propto h^2,
\end{equation}
where we have used $\Pi\propto p/\rho \propto h^2$.  

The height-integrated stress at the ISCO enters the solution as a
boundary condition when integrating the energy equation
(\ref{eq:energy}). It scales as
\begin{equation}\label{eq:stressisco}
\frac{\shear t^{\hat{r}\hat{\phi}} h }{\dot{M}}
\propto \frac{\alpha p h}{\rho u^r h} 
\propto \alpha h.
\end{equation}
In the first step, we used the $\alpha$-viscosity prescription
(\ref{eq:alphastress}). In the second step, we identified the ISCO with
the sonic point of the disk (c.f. \S\ref{sec:bc}), so $u^r\propto c_s \propto
\sqrt{p/\rho} \propto h$ (by \ref{eq:hydroeq}).

Equations (\ref{eq:compress})-(\ref{eq:stressisco}) give the scaling
of compression, advection, and boundary stresses with $h$ and
$\alpha$.  In the NT limit, $h\rightarrow 0$, all three terms vanish.
Under the weaker assumption $h \ll \alpha$, compression and advection
are small but the stress at the ISCO is important.  So we obtain a
self-consistent generalization of the NT model by ignoring advection
and compression but including the boundary stress at the ISCO, when
$h\ll \alpha$.

Dropping advection and compression terms from the energy equation
(\ref{eq:energy}), we have
\begin{equation}\label{eq:energy2}
\frac{d q^z}{dz} = -\shear t^{\hat{r}\hat{\phi}},
\end{equation}
which says energy generated by shear stresses is immediately radiated
away.  Height integrating gives (\ref{eq:fluxstress}).

Rewriting the stress at the ISCO (\ref{eq:stressisco}) as a radiant flux
using (\ref{eq:fluxstress}) fixes the boundary term in the disk solution
(\ref{eq:PTintegral}):
\begin{equation}
C = \left[
\alpha h \gamma \left(E-\Omega L \right)\frac{\abr}{r}
\right]_{0}.\label{eq:intconst}
\end{equation}
This reduces to the NT choice $C=0$ in the razor thin limit $h
\rightarrow 0$.  However in general the flux and stress at the ISCO
will not be zero.

\bibliographystyle{mnras}
\bibliography{ms}

\end{document}